# Coupling single-molecules to DNA-based optical antennas with position and orientation control

*Aleksandra K. Adamczyk[1], Fangjia Zhu[1], Daniel Schäfer[2], Yuya Kanehira[3], Sergio Kogikoski Jr[3], Ilko Bald[3], Sebastian Schlücker[2], Karol Kołątaj[1,6*], Fernando D. Stefani[4,5*], and Guillermo P. Acuna[1,6*]*

[1] Department of Physics, University of Fribourg, Chemin du Musée 3, Fribourg CH-1700, Switzerland.

[2] Department of Chemistry and Center of Nanointegration Duisburg-Essen (CENIDE) & Center of Medical Biotechnology (ZMB), University of Duisburg-Essen, Essen, Germany.

[3] Institute of Chemistry, University of Potsdam, Karl-Liebknecht-Str. 24-25, 14476 Potsdam, Germany.

[4] Centro de Investigaciones en Bionanociencias (CIBION), Consejo Nacional de Investigaciones Científicas y Técnicas (CONICET), Godoy Cruz 2390, C1425FQD Ciudad Autónoma de Buenos Aires, Argentina.

[5] Departamento de Física, Facultad de Ciencias Exactas y Naturales, Universidad de Buenos Aires, Güiraldes 2620, C1428EHA Ciudad Autónoma de Buenos Aires, Argentina.

[6] Swiss National Center for Competence in Research (NCCR) Bio-inspired Materials, University of Fribourg, Chemin des Verdiers 4, CH-1700 Fribourg, Switzerland.

**Corresponding Authors**

*E-mail: karol.kolataj@unifr.ch, fernando.stefani@df.uba.ar, guillermo.acuna@unifr.ch

**Abstract**

Optical antennas have been extensively employed to manipulate the photophysical properties of single photon emitters. Coupling between an emitter and a given resonant mode of an optical antenna depends mainly on three parameters: spectral overlap, relative distance, and relative orientation between the emitter´s transition dipole moment and the antenna. While the first two have been already extensively demonstrated, achieving full coupling control remains unexplored due to the challenges in manipulating at the same time both the position and orientation of single molecules. Here, we use the DNA origami technique to assemble a dimer optical antenna and position a single fluorescent molecule at the antenna gap with controlled orientation, predominately parallel or perpendicular to the antenna's main axis. We study the coupling for both conditions through fluorescence measurements correlated with scanning electron microscopy images, revealing a 5-fold higher average fluorescence intensity when the emitter is aligned with the antenna´s main axis and a maximum fluorescence enhancement of ~ 1400-fold. A comparison to realistic numerical simulations suggests that the observed distribution of fluorescence enhancement arises from small variations in emitter orientation and gap size. This work establishes DNA origami as a versatile platform to fully control the coupling between emitters and optical antennas, trailblazing the way for self-assembled nanophotonic devices with optimized and more homogenous performance.

**Introduction**

Optical antennas[1] (OAs) based on metallic or dielectric nanoparticles (NPs) are one of the basic components of nanophotonic devices, acting as transducers between propagating light and localized fields, enabling a remarkable enhancement of light–matter interactions at the nanoscale[2]. In particular, OAs have been widely demonstrated to control and manipulate the photophysical properties of single photon emitters such as organic fluorophores and quantum dots[3]. The interaction between a photon emitter and a given resonant mode of an OA is mainly governed by their spectral overlap, relative position, and relative orientation[4]. The spectral overlap can be tuned by the choice of OA design and emitter[5]. First studies of the position dependent coupling of an emitter and an OA were done with atomic force microscopy (AFM) techniques[6–9]. While these pioneering measurements provided valuable insight, those AFM-based techniques are challenging to implement for more complex OA geometries involving two or more NPs[10]. Furthermore, they are hardly suited for the fabrication of antenna-emitter systems with controlled relative position, and they provide no control over the emitter orientation.

The advent of the DNA origami technique[11] enabled the bottom-up self-assembly of colloidal metallic[12] and recently dielectric[13,14] NPs together with organic dyes and quantum dots with nanometric positional precision and stoichiometric control[15–17]. In this way, the DNA origami technique was exploited to fabricate diverse OAs with single emitters placed at specific locations to enhance the fluorescence intensity[18,19] and photostability[20], and to direct or tune the emission[21–24]. These DNA origami based OAs can very precisely manipulate the photophysical properties of single photon emitters located at the hotspot as demonstrated for example by maximum values of fluorescence enhancement (FE) reaching three orders of magnitude[25] and forward to backward directivities over 10 dB[22]. Despite these impressive values, the overall performance of these OAs remains rather inhomogeneous with a significant dispersion. This limitation adversely affects the efficiency and reproducibility of OAs and their implementation beyond fundamental research for quantitative applications[26].

We hypothesize that an important factor behind the OA´s inhomogeneous overall performance in the cases reported before lies in the fact that the emitter´s transition dipole orientation was not controlled. Therefore, the OA-emitter coupling can vary from virtually suppressed to its maximum value depending on their relative orientation. However, incorporating molecules in a nanodevice with controlled position, stoichiometry and orientation has remained an open challenge. Recently, we have demonstrated the first steps towards controlling the orientation of single fluorophores in DNA origami structures[27]. Briefly, by engineering both the fluorophore link to DNA and the local environment, the transition dipole moment of Cy3 and Cy5 molecules can be oriented predominantly parallel or perpendicular to the host double-stranded (ds) DNA helix. Analogous results were independently obtained, reinforcing the robustness of this method[28]. Here, we study the use of this approach to achieve full control of the coupling between a fluorophore and an OA.

**Results**

A sketch of the DNA origami employed to self-assemble the OAs and control both the position and orientation of a single emitter is included in Figure 1a. It consists of a two-layer 12-helix rectangular structure with dimensions of approximately 180 nm × 20 nm × 5 nm (length × width × height) with a "mast" at the center (not shown). The DNA origami structure is modified with single stranded (ss-) DNA handles consisting of a A18 sequence to incorporate, through DNA hybridization, two 60 nm gold NPs previously functionalized with the complementary sequence T18 (further details on the DNA origami design are given in the Materials and Methods section and in Figure S1). We estimate the resulting interparticle gap $g = (10 \pm 4)$ nm. Between the Au NPs, the DNA origami hosts a single Cy5 molecule oriented either predominantly aligned with the main dimer axis (sample ∥) or perpendicular to it (sample ⊥), see zoom in Figure 1a. This is achieved by leaving no (∥) or 8 (⊥) adjacent bases unpaired from the DNA origami scaffold so that the fluorophore's orientation is adjusted without changing its position. The OAs were then purified by gel electrophoresis and imaged by transmission electron microscopy

(TEM), confirming the correct self-assembly of the structures (Figure 1b). It is worth noting that commercial Au NPs tend to differ in shape from their nominal spherical description. In addition, a control sample was fabricated using 'bare' DNA origami structures with a single Cy5 without NPs.

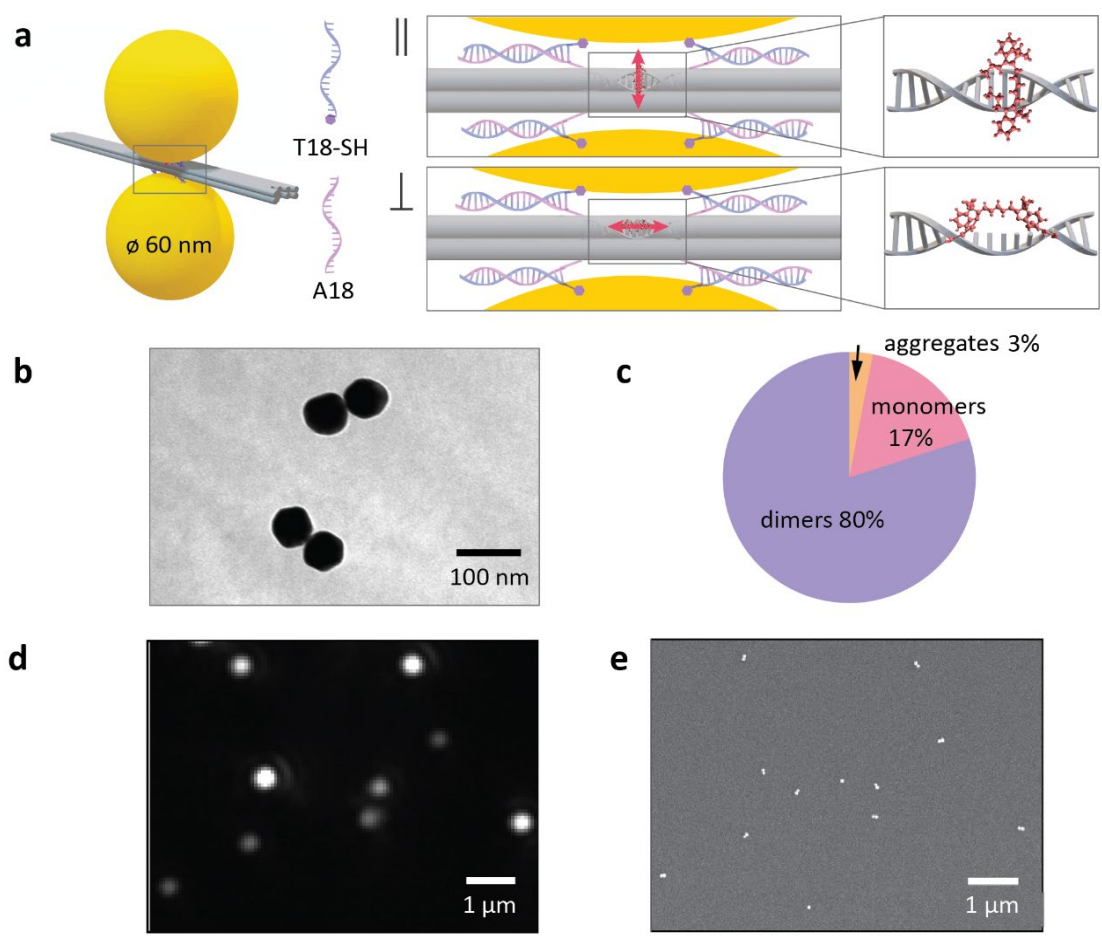

*Figure 1.* Dimer OAs with an oriented single dye at the hotspot. (a) Sketch of the DNA origami host structure and the Au NPs employed. The zoom highlights the gap where a doubly linked fluorophore (symbolized by a red double arrow) was positioned either perpendicular or parallel to the two 60 nm Au NPs forming the OA dimer. (b) TEM image of two OA dimers. (c) Pie chart illustrating the yield of OAs on the sample surface. (d) An exemplary image obtained with widefield fluorescence measurements. Spots of different intensities can be already

*detected. (e) Corresponding SEM image of the identical area, utilized for colocalization with (d) to confirm that the signal originates from dimer OAs.*

The OAs and control structures were immobilized on glass coverslips at a surface density suited for single molecule fluorescence measurements (see Supplementary Note 1: Sample Preparation and Surface Immobilization for details). The emission of individual OAs and control structures were measured in a wide-field microscope under circularly polarized excitation. Subsequently, the samples containing the OAs were dried and transferred to a scanning electron microscope (SEM). As shown in the examples of Figures 1d and 1e, there is a high correlation between the detected fluorescence spots and dimer OAs imaged by SEM. The sample also contains a minor fraction of NP monomers and a small number of dimer OAs that lack fluorescence signal. These combined SEM and fluorescence measurements allowed us to quantify the yield of functional OAs (Figure 1c) and analyze solely the fluorescence arising from dimer structures. The yield of dimer structures was 80%, confirming the high efficiency of the DNA origami OA self-assembly process.

Figure 2a shows a comparison of exemplary fluorescence intensity transients originating from single fluorophores oriented parallel and perpendicularly to the main axis of the OA dimer, and from a single fluorophore in the control sample. Only traces showing a single-step photobleaching were considered to assure that the measurement corresponded to a single molecule. Molecules aligned with the main axis of the OA tend to show a significantly higher intensity. This is a direct consequence of two main effects. First, a more efficient excitation because the absorption dipole of the fluorophore is aligned with the gap field of the main resonant antenna mode, as schematically shown in Figure 2b. Second, the fluorophore´s quantum yield will be significantly more quenched when aligned perpendicular to the main resonant antenna mode[29]. This can be rationalized for example by considering the electric dipoles induced by the fluorophore on the NPs, that will interact destructively with the fluorophore´s emission dipole, leading to a reduced emission into the far field[30,31].

We measured fluorescence intensity transients of 404, 479 and 140 structures corresponding to samples ∥ , ⊥ and the control, respectively. Figure 2c summarizes the main results, showing the distribution of fluorescence enhancement for samples ∥ and ⊥, computed as the average intensity of each trace normalized to the average fluorescence intensity of all molecules in the control sample (see Figure S2). While both samples exhibit a rather broad dispersion of enhancement factors, the values obtained in the ∥ sample are significantly higher. The distributions of enhancement factors are well fit with a log-normal distribution[32] (solid lines in Figure 2c), retrieving the following mean fluorescence enhancement ($\mu$) and standard error ($SE$): $\mu_\parallel$ = 245.3 , $SE_\parallel$ = 13.4 , $\mu_\perp$ = 46.1 and $SE_\perp$ = 3.0. Based on these results, we can conclude that by controlling the orientation of the fluorophore without affecting its position the fluorescence signal can be increased 5-fold in average. Notably, it is worth highlighting that a maximum enhancement factor of ~ 1400 was achieved for sample ∥. This is a remarkably high value for OA dimers based on 60 nm Au NPs at an average gap of 10 nm. This value translates into a fluorescence enhancement figure of merit (defined as FE × $\phi_0$) of ~ 400 (considering the intrinsic quantum yield of Cy5 $\phi_0$ ~ 0.3), which is the highest one reported to date[33] in the visible range.

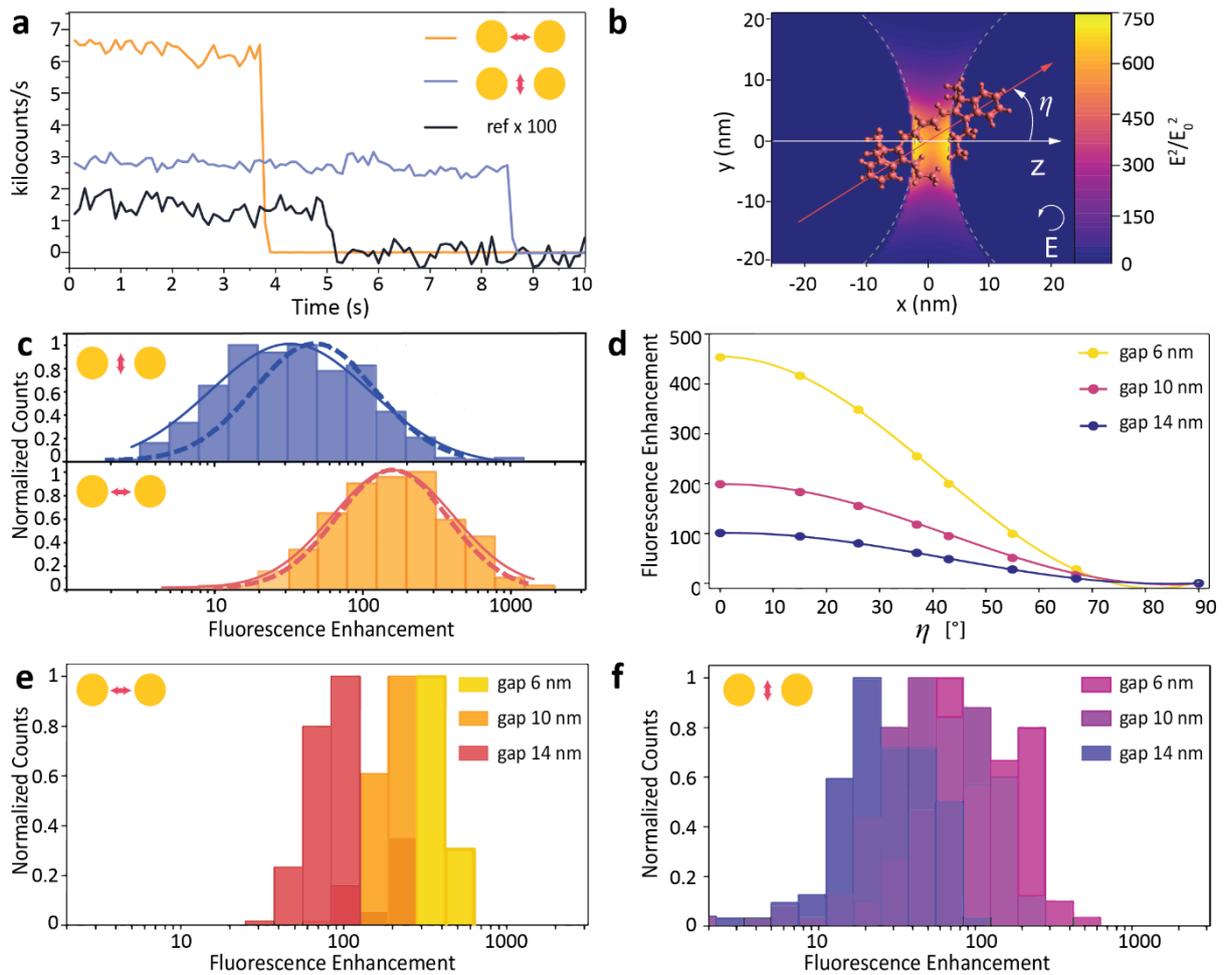

*Figure 2.* Fluorescence enhancement of OA´s with oriented fluorophores at the gap: measurements and simulations. (a) Examples of fluorescence intensity traces for an OA with the fluorophore oriented parallel and perpendicular to the main axis. For comparison a reference transient from a DNA origami containing a single Cy5 dye but no Au NPs is included. (b) Simulations of the electric field intensity enhancement for a 6 nm gap at an excitation wavelength of 640 nm with circularly polarized light. The orientation of the electric field at the gap is predominantly in line with the dimer axis (not shown). The coordinate system employed to describe the orientation of single fluorophores in the antenna is included. (c) Fluorescence enhancement histogram plot for 60 nm Au OAs dimers with a single Cy5 dye located at the hotspot oriented predominantly parallel (orange) or perpendicular (blue) to the main dimer

*axis. The solid lines represent a log-normal distribution fit whereas the dashed lines correspond to numerical simulations. (d) Simulated fluorescence enhancement dependent on the fluorophore orientation for three different systems with interparticle distances equal to 6 nm, 10 nm, and 14 nm, respectively. The simulated fluorescence enhancement distributions for samples ∥ (e) and ⊥ (f), considering gaps of 6, 10, and 14 nm and the previously measured angular distributions[27].*

The results in Figure 2c clearly show that the molecular orientation can be controlled leading to a strong effect on the coupling to the OA. However, the distributions of observed enhancement factors are relatively broad, as a result of several factors. The excitation rate might be inhomogeneous if for example, the OAs have a distribution of out-of-plane orientations. However, this is unlikely since due to the dimensions of the NPs and the DNA origami employed together with the binding scheme, OAs will predominantly lie flat on the glass coverslip surface. The dispersion of size and shape of the NPs might also affect the homogeneity of results (see for example Figure 1b). To test this point, we have repeated the experiments using "super-spherical" 50 nm Au NPs[34], see Figure S3. These OAs display an overall lower enhancement as expected from the smaller NPs size (Figure S5). Nonetheless, the observed distributions of fluorescence enhancement were equivalently broad to the ones obtained with the commercial NPs. We therefore conclude that the dispersion of NPs size and shape does not play a crucial role in the inhomogeneity of coupling observed.

Other factors that can lead to a dispersion of fluorescence enhancement are variations of molecular orientation and gap size. To analyze this we performed numerical calculations to determine the expected fluorescence enhancement for a range of orientations and gaps. Figures 2d and S4 summarize the results, which show that a higher fluorescence enhancement is expected for smaller gaps and orientations close to the main axis of the OA. For the perpendicular orientation no enhancement is expected regardless of the gap size. The observed fluorescence enhancement for the ⊥ sample, with maximum values even reaching several 100 fold, indicates that even for this configuration the molecular dipole orientation has a significant component along the main axis of the OA. This observation, together with the broad distribution of enhancement factors exposes the limitations of the approach to fix the molecular orientation, which has been characterized in a previous work[27].

As to the gap size, standard deviations from 1 to 2.7 nm have been already observed for gaps in different DNA origami based OAs[18,35–37]. It is worth noting that those reported values were estimated from TEM measurements on dried samples. However, our measurements were performed in liquid and therefore a higher dispersion in gap sizes could be expected. To obtain a high yield of Au OAs, we decided to use two binding configurations between DNA origami handles and Au NPs, namely the shear and zipper configuration[38], see Figure S1. Therefore, considering both the number and the position of the binding sites employed to incorporate the 60 nm Au NPs, we estimate that the gap size lies between 6 and 14 nm for zipper and shear geometry of binding, respectively.

To study the influence of gap size and molecular orientation variations on the fluorescence enhancement, we carried out numerical simulations to calculate the expected distributions of fluorescence enhancement for three gap sizes 6, 10 and 14 nm (zipper, mixed, and shear binding configurations), considering the molecular angular distributions previously determined for each configuration (Further simulation details are included in the Supplementary Note 3). Figures 2e and 2f show the simulated fluorescence enhancement distributions for samples ∥ and ⊥, considering gaps of 6, 10, and 14 nm. In Figure 2c, the simulated distributions for the three gaps are summed up and presented as bold dashed lines to compare to the experimental results. The numerical simulations performed with the input of the measured three-dimensional angular distributions of the fluorophores agree well with the experiments.

**Discussion**

In conclusion, this work demonstrates how DNA nanotechnology can be used to control the coupling of a single molecule to the resonant mode of an OA by simultaneously controlling the position and the orientation of the molecule relative to the OAs. The strategy applied involves a double link of the fluorophore to the DNA double helix with different number of unpaired bases to modulate the local environment. In this way, a remarkable fivefold increase in the average fluorescence intensity is achieved, along with an unprecedented maximum enhancement of 1400-fold. Based on control experiments using super-spherical NPs and by comparing the distributions of enhancement factors to realistic simulations, we can conclude that dispersion of results arises from variations in molecular orientations and gap sizes. Looking ahead, it is worth exploring more rigid OA designs aiming to deliver more uniform gap sizes as well as new strategies for a more precise control of molecular orientation. Finally, our results firmly establish the DNA origami technique as a versatile and comprehensive nanofabrication platform to control the position and orientation of different species such as organic dyes and NPs to fully manipulate their interaction. This assembly precision holds promise for achieving OAs with maximum efficiency and reproducibility, a crucial step for advancing their implementation in quantitative nanophotonic applications.

**Materials and methods**

DNA origami design and folding,

The square-lattice DNA origami structure was designed using CaDNAno[39], and it is accessible at nanobase.org[40]. The 7249 bases scaffold (M13mp18, Bayou Biolabs LLC) and 186 synthetic staple strands were folded in 1× TAE (Alfa Aesar, #J63931) and 12 mM $MgCl_2$ (Alfa Aesar, #J61014) using a 1:10 scaffold/staples ratio and 1:100 for modified staples. Unmodified staples as well as staples internally modified with Cy5 (iCy5, product #1476) were purchased from

Integrated DNA Technologies, INC, and biotin-functionalized were purchased from Biomers GmbH. The scaffold and staples mix were initially heated and held at 70 °C for 5 min before being cooled to 22 °C using a 20 min/1 °C linear ramp. A 1% agarose gel electrophoresis (Agarose LE, Biozym Scientific GmbH) was used as a purification procedure to remove the excess of staple strands (1× TAE 12 mM $MgCl_2$ running buffer, 4 V/cm, 1.5 h) while being cooled in an ice water bath. After electrophoresis, the bands in the gel containing the DNA origami structure were cut out and squeezed with a glass slide covered in parafilm to extract the purified DNA origami structures. The final concentration of the DNA origami structures was determined on a Nanodrop 2000 spectrophotometer (Thermo Fisher Scientific). It is important to note that the DNA structures were not stained.

Nanoantenna self-assembly

To functionalize commercially available, citrate-capped 60 nm AuNPs (BBI) with thiolated ssDNA (T8 and T18 in 4:1 ratio, Ella Biotech GmbH), disulfide bonds are cleaved with tris (2-carboxyethyl) phosphine at room temperature for 1 h. Then, the cleaved oligonucleotides are mixed with AuNPs and left in a freezer at −20 °C for 3 h. After being thawed, AuNPs are purified from the aggregates and an excess of DNA strands by gel electrophoresis (1× TAE 12 mM $MgCl_2$ running buffer, 4 V/cm, 1.5 h, on ice water bath). The AuNPs are recovered by cutting the band with the highest electromobility.

Purified origamis and freshly functionalized AuNPs are mixed in a 1:10 molar ratio, and 600 mM NaCl is added. Then, this mixture is left overnight in room temperature to allow assembly. Dimers are purified from the rest by gel electrophoresis. This solution is loaded into 1% agarose gel and separated for 4 h on ice water bath. Dimers are recovered by cutting the respective band and then immobilized on a glass substrate.

Sample Preparation and Surface Immobilization

For immobilization of the structures, glass coverslips were first rinsed with water and then cleaned in a UV cleaning system (PSD Pro System, Novascan Technologies, USA) followed by incubation in 3 M KOH for 5 min. After rinsing with 1× PBS (Alfa Aesar, #J75889), the surface was passivated with BSA biotin (0.5 mg/mL, Sigma-Aldrich Chemie GmbH, #a8549) and neutrAvidin (0.5 mg/mL, Fisher Scientific AG, #10443985), both sequentially incubated for 30 min and washed with 1× PBS buffer. Then, an additional wash using the origami buffer (1× TAE 12 mM $MgCl_2$) prepared the surface for the incubation, and 15 pM of DNA origami structure was immobilized via biotin binding to the functionalized surface. For nanoantenna immobilization, after incubation with neutravidin the surface was passivated with 500 nM of ss-DNA modified with biotin (biotin-A18 Ella Biotech GmbH), for 15 min and then washed with origami buffer (1× TAE 12 mM $MgCl_2$). After 15 min of incubation in case of origami, and 30 min in case of antennas, the sample was washed, and the buffer was exchanged to a buffer containing coupled glucose oxidase and catalase (GODCAT) in a buffer containing 1 mM glucose in order to increase the photostability of the fluorophores.

Widefield measurements

Measurements were performed on a home-build TIRF wide-field microscope built on an inverted Olympus IX83 body. For excitation, a 640 nm laser (Laserquantum, gem640) is used. Spectral clean-up of the laser's emission is performed through the filters (ZET532/640, Chroma Technology Corporation). Dichroic Mirrors (DM) are used to join the path and πShaper (VIS Flat Top Beam Shaper | πShaper 6_6_VIS) is used providing a flat beam profile suitable for scenarios requiring uniform illumination. Linear polarizer and λ/4 plates (B-Halle, # RAC 3.4.10) were included in the path for circular excitation polarization. The laser light is then focused by two lenses (AC508-100-A-ML and AC254-030-A-ML) into the back focal plane of the UPLAPO100xOHR (1.5 NA, Olympus) objective. A Laser Dual Band Set (ET- 532/640 nm, DM3) is used to reflect the excitation towards the sample and transmit the emission to reach the CMOS camera (C14440 ORCA-Fusion, Hamamatsu). Data acquisition is performed using the open source microscopy imageJ software Micro-Manager[41].

Data availability

The data underlying the results presented in this paper is available from the corresponding author upon reasonable request.

**Acknowledgements**

This project has received funding from the European Union's Horizon 2020 research and innovation programme under the Marie Skłodowska-Curie grant agreement No 860914. F.D.S. acknowledges the support of the Max Planck Society and the Alexander von Humboldt Foundation. This work has been funded by Consejo Nacional de Investigaciones Científicas y Técnicas (CONICET) and Agencia Nacional de Promoción Científica y Tecnológica (ANPCYT), project PICT-2017-0870. G.P.A. acknowledges support from the Swiss National Science Foundation (200021_184687) and through the National Center of Competence in Research Bio-Inspired Materials (NCCR, 51NF40_182881), the European Union Program HORIZON-Pathfinder-Open: 3D-BRICKS, grant Agreement 101099125. FDS acknowledges the Alexander von Humboldt foundation for their constant support. IB and YK acknowledge support by the European Research Council (ERC; consolidator Grant No. 772752).

**Conflict of interest**

The authors declare no competing interests.